\documentclass{elsarticle}

\usepackage{amsmath}
\usepackage{amssymb}
\usepackage{amsfonts}
\usepackage[colorlinks=false,hypertexnames=false]{hyperref}
\usepackage{hypcap}
\usepackage{multirow}
\usepackage{placeins}
\usepackage[table]{xcolor}
\usepackage[normalem]{ulem}
\usepackage{comment}
\hypersetup{pdfborder={0 0 0},pdftitle={}}
\makeatletter
\renewenvironment{figure}[1][\fps@figure]{
\edef\@tempa{\noexpand\@float{figure}[#1]}
\@tempa\capstart
}{
\end@float
}
\makeatother
                                \hypersetup{pdftitle={GTV learning}}

\usepackage{geometry}
\usepackage{setspace}
\emergencystretch=\maxdimen
\author[1,3]{Thibault Marin\fnref{fn1}}
\author[1,3]{Yue Zhuo\fnref{fn1}}
\author[1,3]{Rita Maria Lahoud}
\author[1,3]{Fei Tian}
\author[1,3]{Xiaoyue Ma}
\author[1,3]{Fangxu Xing}
\author[1,2,3]{Maryam Moteabbed}
\author[1,3]{Xiaofeng Liu}
\author[1,3]{Kira Grogg}
\author[2,3]{Nadya Shusharina}
\author[1,3]{Jonghye Woo}
\author[1,3]{Chao Ma}
\author[1,2,3]{Yen-Lin E. Chen}
\author[1,3]{Georges El~Fakhri\corref{cor1}}
\fntext[fn1] {Equal contribution}
\cortext[cor1]{Corresponding author, elfakhri.georges@mgh.harvard.edu}
\address[1]{Gordon Center for Medical Imaging, Department of Radiology, Massachusetts General Hospital, Boston MA, 02114, United States of America}
\address[2]{Department of Radiation Oncology, Massachusetts General Hospital, Boston MA, 02114, United States of America}
\address[3]{Harvard Medical School, Boston MA, 02115, United States of America}
\date{May 12, 2021}
\title{Modeling inter and intra observer variability of GTV contouring using deep learning}
\hypersetup{
 pdfauthor={},
 pdftitle={Modeling inter and intra observer variability of GTV contouring using deep learning},
 pdfkeywords={},
 pdfsubject={},
 pdfcreator={Emacs 27.1 (Org mode 9.4.3)}, 
 pdflang={English}}
\begin{document}

\begin{frontmatter}
\begin{abstract}

\emph{Background and purpose}: The delineation of the gross tumor volume (GTV) is a
critical step for radiation therapy treatment planning.  The delineation
procedure is typically performed manually which exposes two major issues: cost
and reproducibility. Delineation is a time-consuming process that is subject to
inter- and intra-observer variability.  While methods have been proposed to
predict GTV contours, typical approaches ignore variability and therefore fail
to utilize the valuable confidence information offered by multiple contours.

\emph{Materials and methods}: In this work we propose an automatic GTV contouring
method for soft-tissue sarcomas from X-ray computed tomography (CT) images,
using deep learning by integrating inter- and intra-observer variability in the
learned model. Sixty-eight patients with soft tissue and bone sarcomas were
considered in this evaluation, all underwent pre-operative CT imaging used to
perform GTV delineation.  Four radiation oncologists and radiologists performed
three contouring trials each for all patients.  We quantify variability by
defining confidence levels based on the frequency of inclusion of a given voxel
into the GTV and use a deep convolutional neural network to learn GTV confidence
maps.

\emph{Results}: Results were compared to confidence maps from the four readers as
well as ground-truth consensus contours established jointly by all readers.  The
resulting continuous Dice score between predicted and true confidence maps was
87\% and the Hausdorff distance was 14 mm.

\emph{Conclusion}: Results demonstrate the ability of the proposed method to predict
accurate contours while utilizing variability and as such it can be used to
improve clinical workflow.

\end{abstract}


\end{frontmatter}

\section*{Introduction}
\label{sec:org5444533}

While radiotherapy (RT) is one of the main treatment options for soft tissue and
bone sarcoma, achieving consistently good treatment outcomes for high-risk
sarcomas remains challenging.  There is now mature evidence supporting the use
of preoperative radiation therapy in sarcomas where margins are anticipated to
be close or at high risk of local recurrence~\cite{Wang2015a}.  Benefits
of radiation includes visibility of tumor allowing consistent target volume
definition, prevention of seeding, reducing necessary surgical margins, reducing
the volume of radiated tissue, and thereby decreasing late
toxicities~\cite{Wang2015a}.  Accurate GTV definition is the first step
to effective radiotherapy, but depending on the quality of the CT simulation
(use of contrast, CT resolution, ability to fuse with MRI image), GTV definition
can be time consuming with varying accuracy \cite{Wang2011,Anderson2014,Ng2018}.
Improving automation of GTV is the first step to automate target definition and
reducing through-put time for initiating preoperative radiation.

In recent years, deep learning has emerged as a powerful tool for medical image
analysis.  The advent of convolutional neural networks coupled with the
availability of fast computing hardware has enabled the development of high
performance algorithms for a variety of medical tasks, including disease
detection and image segmentation.  Deep neural networks (DNNs) are able to
approach and even sometimes surpass human performance in medical analysis tasks.
In the context of automatic GTV delineation, DNNs have been shown to predict the
tumor volume from a single reader with reasonable accuracy using CT
\cite{Cardenas2018,Men2017,Li2018}, PET/CT
\cite{Huang2018a,Guo2019,Jin2019,Moe2019,Ikushima2017}, and MRI
\cite{Huang2018,Hermessi2019,Lin2019a}.  These methods rely on ground-truth
tumor contours obtained by a single human reader and used train the neural
network.  Therefore, they are dependent on the specific contour used as
ground-truth and do not account for the variability between contouring sessions
and readers.  This variability can also be seen as an indicator of the
confidence in the GTV delineation, and therefore can bring valuable
probabilistic information if properly modeled.  Neural networks can be designed
to predict probabilities, for instance the probabilistic
U-Net~\cite{Kohl2018} which uses a variational network to generate a
distribution of plausible segmentations.

In this work, we propose a method to automatically predict the GTV from CT
images using a convolutional neural network.  The proposed method models intra-
and inter-reader variability by learning GTV confidence maps constructed from
multiple GTV contours and representing a confidence level for each image pixel.
The convolutional neural network builds upon the widely used
U-Net~\cite{Ronneberger2015}, adapted to learn and predict confidence
maps rather than binary masks.  Learning of confidence maps is modeled as an
ordered multi-class classification task and training is driven by a ranking
loss~\cite{Cheng2008}.  The proposed method is validated by comparing
predicted confidence maps to ground-truth confidence maps obtained from human
readers as well as a consensus GTV established by all the readers in a joint
contouring session.

\section*{Methods and Materials}
\label{sec:org24955c0}

\subsection*{Dataset}
\label{sec:org572d109}

Seventeen patients with soft-tissue and bone sarcomas underwent pre-operative
non-contrast CT at Massachusetts General Hospital (MGH), in accordance with the
Institutional Review Board (IRB).  The dataset consisted of 11 patients with
soft-tissue sarcoma (including 5 extremity tumors) and 6 patients with bone
sarcomas (chordomas).

The dataset was augmented by using the public soft-tissue sarcoma dataset from
The Cancer Imaging Archive (TCIA)~\cite{Vallieres2015}, which contains 51
patients imaged by CT with soft-tissue sarcoma confirmed by histology.  The
total number of patients used in this work was 68.

In order to evaluate the performance of the proposed neural network, a
representative subset of the patients was excluded from the training procedure
and reserved for evaluation.  This evaluation dataset was selected to cover
different types of tumors and was composed of two bone sarcomas (chordomas), and
six soft-tissue sarcomas: one pelvic and five in extremities.

\subsection*{GTV delineation}
\label{sec:org74b5835}

\begin{figure*}
  \centering
  \includegraphics{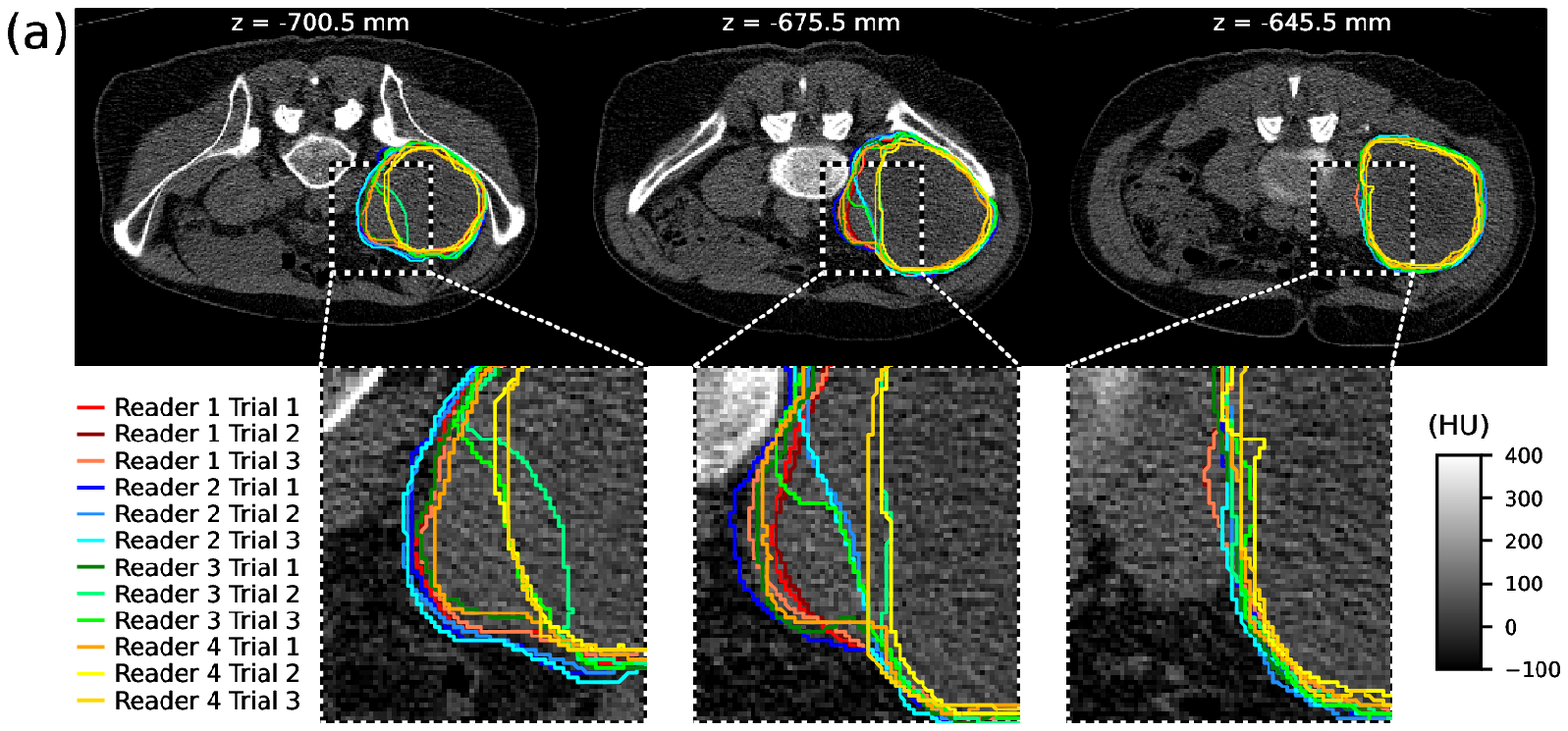}
  \\
  \includegraphics{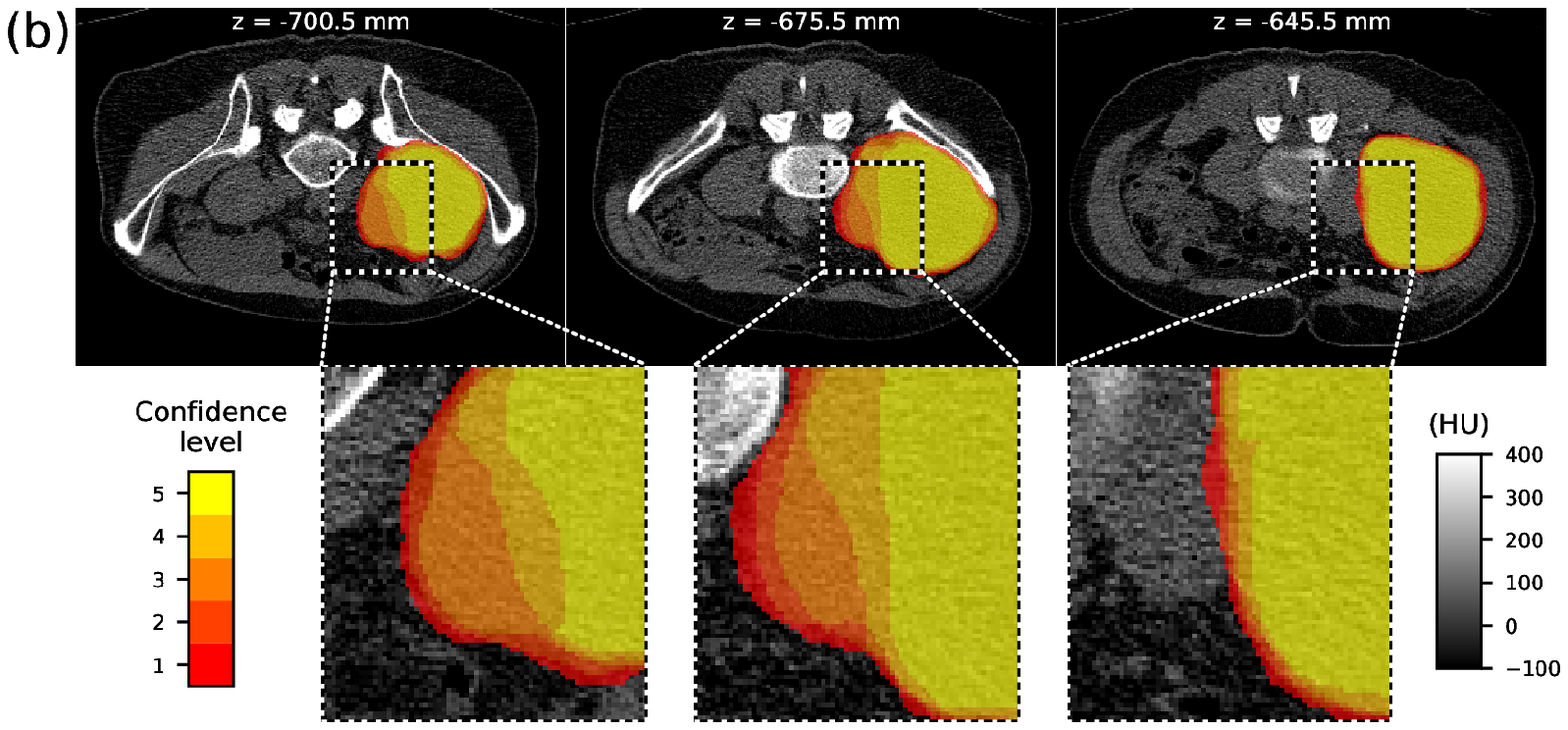}
  \caption{\label{fig-contours}GTV contours and confidence maps. (a) GTV
    contours from 4 readers with 3 trials each drawn on CT images at different
    slices.  The average Dice coefficient between pairs of is 88\% (Dice scores
    range from 78\% to 97\%).  (b) Confidence map calculated from the contours
    shown above.  The color bar shown on the bottom left indicates the level of
    confidence in including each pixel in the GTV (from 0 to 5).}
\end{figure*}

For each patient, 4 readers (radiologists and radiation oncologists) performed
GTV delineation using the MIMVista software (MIM Software Inc., Cleveland, OH).
GTV delineation was performed slice by slice and repeated three times by each
reader, in a randomized order to avoid recall bias.  Due to missing data for
some patients, the total number of GTV contouring trials for each patient varied
between 6 and 12.  Example contours for a representative patient are shown in
Figure~\ref{fig-contours}(a), demonstrating the typical variability in
GTV contours.

In order to model the variability, the contours were combined into a discretized
confidence map.  Each contour was first rasterized to generate a binary mask;
masks were then summed and the result was quantized into 6 values (from 0 for
pixels that were never included in a GTV contour to 5 for pixels included in
every available GTV contour).  An example of confidence map overlaid on the
corresponding CT image is shown in Figure~\ref{fig-contours}(b).  Using
discretized maps allows flexibility in the number of contours available for
training.

In addition to the three trials, readers jointly determined a consensus GTV
contour for each patient in a separate session for accuracy evaluation of both
human contours and contours predicted by the network.

\subsection*{Convolutional neural network}
\label{sec:org59c24fc}

We developed a convolutional neural network to predict GTV confidence maps in
order to model inter- and intra-reader variability.  The training task is viewed
as a multi-class segmentation problem, i.e. learning the discrete confidence
level for each pixel, with the important characteristic that classes
(i.e. confidence levels) are ordered.

The network was based on the U-Net~\cite{Ronneberger2015} which has been
extensively used for image segmentation tasks in the recent years
\cite{Li2018,Siddique2020,Sun2020}.  The U-Net architecture is based on the
combination of contracting/expanding paths with skip connections capturing
spatial information.  The contracting path is composed of successive
convolutional layers interleaved with nonlinear activation (ReLU), batch
normalization and pooling layers responsible for enforcing a sparse
representation at the bottleneck level and increasing the receptive field by
downsampling images at each level.  The expanding path is constructed as the
dual of the contracting path, replacing pooling with upsampling and adding skip
connections.  The final layer of the network is a convolutional layer with \(C\)
channels (where \(C\) is the number of output classes).  It predicts a probability
for each class and each pixel.

\begin{figure*}
\centering
\includegraphics{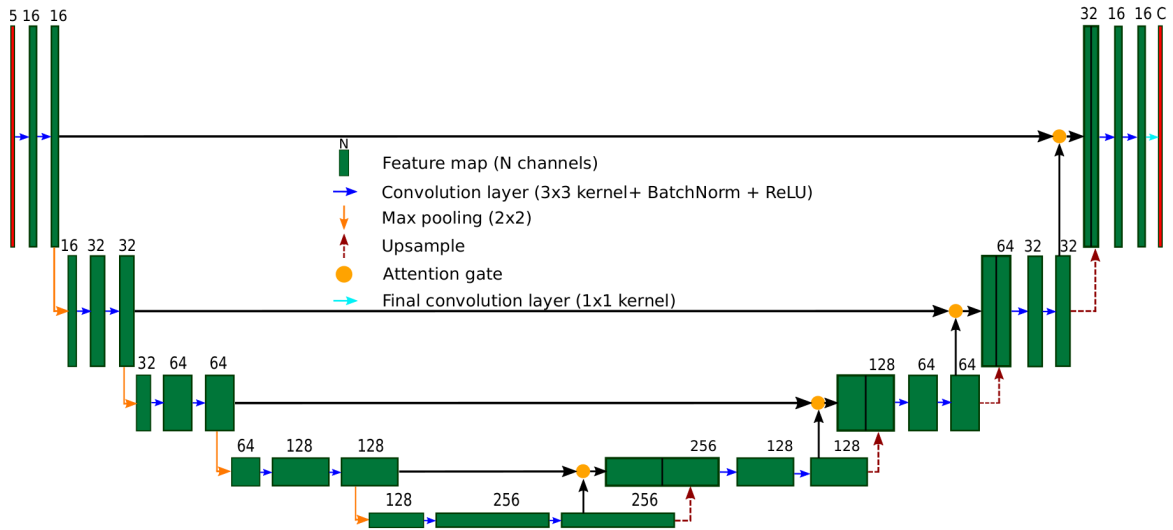}
\caption{\label{fig-unet}Neural network used to predict GTV confidence maps.  Green boxes represent feature maps with the depth indicated above the box. Horizontal blue arrows represent convolution layers, orange arrows represent downsampling by max-pooling, dashed red arrows represent upsampling and orange discs indicate attention gates as described in~\cite{Oktay2018}.}
\end{figure*}

In this work, the U-Net architecture was improved using several techniques
(Figure~\ref{fig-unet}).  First, due to the limited amount of training data
available and to limit the memory requirements, a so-called 2.5D U-Net was used
where images are processed slice by slice while stacking neighboring slices in
the channel dimension of the input images (5 slices were used in this work).
This sliding-window approach enables the use of 3D contextual information
without dramatically increasing the size of the network (and therefore the
amount of labeled data required for training).  Additionally, the network
included attention gates in the skip connections as described in
\cite{Oktay2018}.  The goal of attention-based approaches is to focus the
training on image features directly related to the output class.  By using
self-trained attention maps, the proposed network becomes less sensitive to
class imbalance.

The loss function used to train the network was a modified cross-entropy loss,
which can be expressed as:
\begin{align}
\label{eq-ce}
  \mathcal{L}(\tilde{y}, y) = \sum_j \sum_{c=1}^C
  w_c \, y_j^{(c)} \log(\tilde{y}_j^{(c)}),
\end{align}
where \(C\) is the total number of confidence levels (6 in this work), \(w_c\) is
the class weight for class \(c\) defined by the reciprocal of the class frequency,
\(\tilde{y}\) is the network output, \(y\) is the ground-truth confidence map,
\(y_j^{(c)}\) is the known confidence level for pixel \(j\) and class \(c\), and
\(\tilde{y}_j^{c}\) is the network-generated probability for pixel \(j\) and class
\(c\).  To account for the ordering between confidence levels, network labels are
transformed as described in~\cite{Cheng2008}.  Instead of encoding the
target vector for a pixel \(\left\{y_j^{(c)}\right\}_{c=1,\ldots,C}\) as \((0,
\ldots, 0, 1, 0, \ldots, 0)\) with a probability 1 for the true class index \(k\)
(known as ``one-hot'' encoding), it is expressed as: \((1, \ldots, 1, 1, 0, \ldots,
0)\) where the first \(k\) entries are 1.  This change implies changing the final
activation function from soft-max (traditionally used for classification) to
sigmoid.

Network training was performed using the Adam
optimizer~\cite{Kingma2015}.  Several techniques were used to avoid
over-fitting~\cite{Goodfellow2016}.  Firstly, an \(\ell_2\) penalty on the
network weights was added to the loss.  Secondly, dropout layers were inserted
in the network.  Dropout layers randomly disable a subset of network weights
during training which typically improves the robustness of the network.
Finally, data augmentation was used during training, including image mirroring,
zoom and translation operations.  Data augmentation adds variety to the training
dataset and typically reduces over-fitting.  To select hyper-parameters
(learning rate, dropout rate, \(\ell_2\) weight regularization), a 4-fold cross
validation on the training dataset was performed~\cite{Bergstra2011}.
Complete network parameters are given in Supplementary material~S1.

\subsection*{Evaluation metrics}
\label{sec-metrics}
Several metrics are commonly used to compare binary masks (e.g. Dice, accuracy,
etc.).  In this work, the output of the proposed neural network is a confidence
map instead of a binary mask.  Therefore, we use generalizations of common
metrics to continuous masks as described in~\cite{Taha2015}.  To
calculate the metrics, the generalized true positive (\(\mathrm{TP}\)), false
positive (\(\mathrm{FP}\)), true negative (\(\mathrm{TN}\)) and false negative
(\(\mathrm{FN}\)) counts between a reference mask \(\overline{m}\) and a given mask
\(m\), both continuous are defined as follows:
\begin{align}
  \mathrm{TP} &= \sum_{j=1}^N \min\left(m_j, \overline{m}_j\right),\nonumber\\
  \mathrm{TN} &= \sum_{j=1}^N \min\left(1 - m_j, 1 - \overline{m}_j\right),\nonumber\\
  \mathrm{FP} &= \sum_{j=1}^N \max\left(m_j - \overline{m}_j, 0\right),\nonumber\\
  \mathrm{FN} &= \sum_{j=1}^N \max\left(\overline{m}_j - m_j, 0\right)\nonumber
\end{align}
where \(N\) is the total number of voxels in a region around the tumor.  The
region is based on the bounding box of the tumor, extended on all dimensions by
20\% of the bounding box dimension.  These definitions match the traditional
binary definitions when the masks are binary.  The volumetric metrics used for
evaluation are:
\begin{subequations}
  \label{eq-metrics}
  \begin{align}
    \mathrm{DICE} &= \frac{2 \mathrm{TP}}
                    {2 \mathrm{TP} + \mathrm{FP} + \mathrm{FN}}\label{eq-dice},\\
    \mathrm{Accuracy} &= \frac{\mathrm{TP} + \mathrm{TN}}
                        {\mathrm{N}}\label{eq-acc},\\
    \mathrm{Sensitivity} &= \frac{\mathrm{TP}}
                           {\mathrm{TP} + \mathrm{FN}}\label{eq-sens},\\
    \mathrm{Specificity} &= \frac{\mathrm{TN}}
                           {\mathrm{FP} + \mathrm{TN}}\label{eq-spec}.
  \end{align}
\end{subequations}

Besides the continuous overlap metrics, the 95th percentile Hausdorff distance
was also calculated~\cite{Crum2006}.  Since this metric requires two
binary masks (defined by their contours \(\mathcal{X}\) and \(\mathcal{Y}\)), it can
be calculated after thresholding the confidence maps and can be expressed as:
\begin{align}
\label{eq-hausdorff}
  \mathrm{HD95}(\mathcal{X}, \mathcal{Y}) = P_{95\%}\Big(
  &\left\{\min_{y\in \mathcal{Y}} \left\|x - y\right\|,
    x \in \mathcal{X}\right\}
  \cup \nonumber\\
  &\left\{\min_{x\in \mathcal{X}} \left\|x - y\right\|,
    y \in \mathcal{Y}\right\}
  \Big),
\end{align}
where \(P_{95\%}(.)\) is an operator extracting the 95-th percentile.

Each metric can be computed between the predicted confidence map and the
confidence map obtained by combining GTV contours from all readers and between
predicted or manual contours and the consensus GTV contour established by all
the readers.

\section*{Results}
\label{sec:orge695f04}

\subsection*{Comparison between predicted and true confidence maps}
\label{sec:org0ba6d15}

\begin{figure*}
  \centering
  \includegraphics{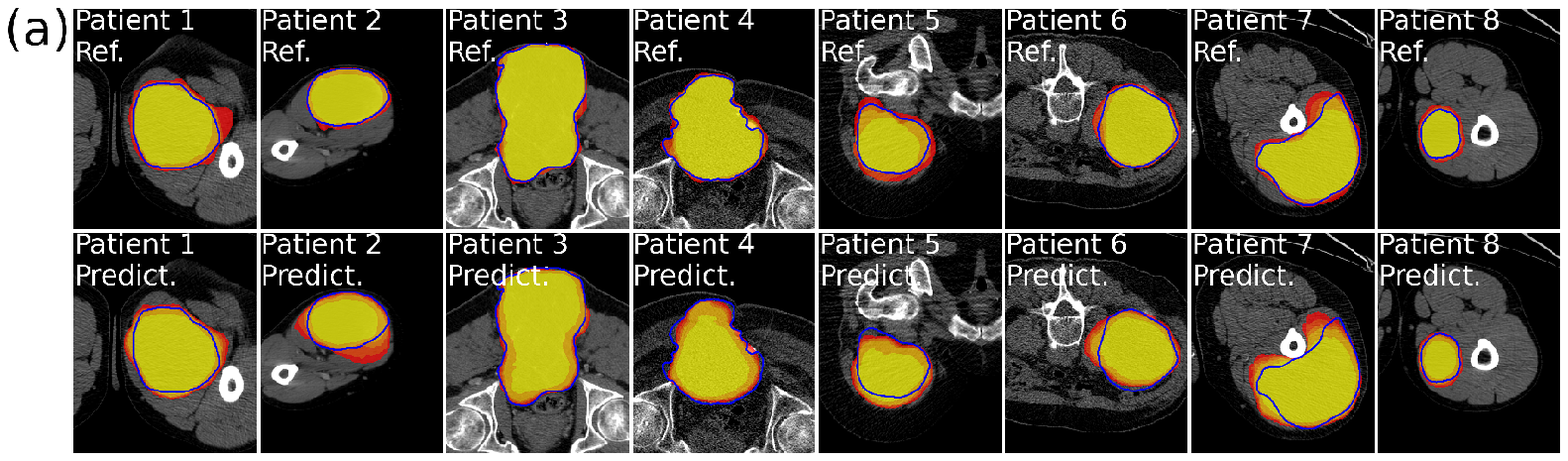}
  \\
  \includegraphics{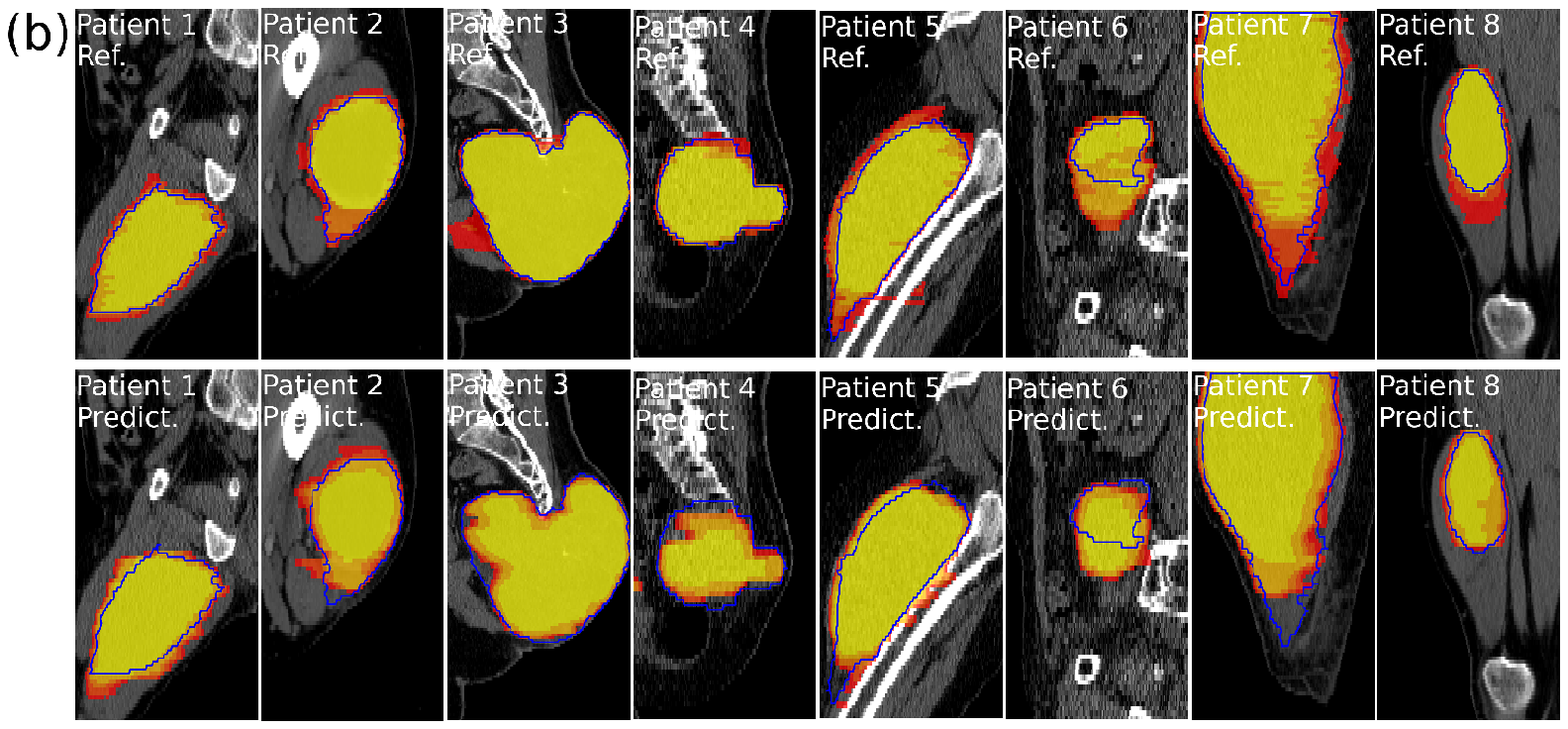}
  \caption{\label{fig-tr-res-z}Confidence map prediction results.  Comparison
    between confidence maps predicted using the proposed deep neural network and
    human confidence maps in (a) axial and (b) sagittal view.  Each column
    corresponds to one patient.  The top row (Ref.)  shows the ground-truth
    confidence map with identical colormaps to Figure~\ref{fig-contours}.  The
    bottom row (Predict.)  shows the output of the proposed deep neural network.
    The blue contour is the consensus GTV jointly determined by all readers.
    The figure demonstrates the good match between predictions and true maps.}
\end{figure*}

The trained network was used to predict confidence maps on the evaluation
dataset.  Predicted confidence maps for eight patients are compared to the
ground-truth human confidence maps in axial and sagittal views in
Figure~\ref{fig-tr-res-z}.  The figure demonstrates a good agreement
between predicted (bottom row) and ground truth maps (top row).  The proposed
deep neural network successfully predicts confidence maps for GTV that are
comparable to confidence maps obtained from multiple readers.

The performance of the proposed network on the evaluation dataset was quantified
using continuous Dice and accuracy metrics between predicted and true confidence
maps.  The proposed method achieves an average continuous Dice coefficient of
86.8\% (\(\pm 5.4 \%\)) and 95\% (\(\pm 1.5 \%\)) accuracy on average, demonstrating a
good agreement between predicted and true confidence maps.

Besides the continuous metrics, a comparison between binary masks obtained by
thresholding confidence maps was performed (see
Table~\ref{tbl-tr-thr-metrics}).  This comparison helps characterize the
performance of the proposed neural network for each confidence level.  Results
suggest a good agreement for all thresholds, with a slight degradation for the
highest threshold, i.e. the highest confidence level.  The threshold confidence
level selected as starting point to draw clinical target volume (CTV) contours
was three, for which a Dice score of 86.7\% (\(\pm 5.4 \%\)) was measured.  For
comparison, the inter-reader variability, measured by calculating the Dice score
between pairs of contours from different readers, was also calculated.  The Dice
score between pairs of human  was around 90.5\% (\(\pm 4.3 \%\)), suggesting that
the level of agreement between predicted and true confidence maps is similar to
the level of agreement expected between human readers.

\begin{table*}[htbp]
\caption{\label{tbl-tr-thr-metrics}Quantitative metrics comparing binary masks obtained by thresholding predicted and ground-truth confidence maps (standard deviation between parentheses).  The column with bold font corresponds to the threshold used as starting point to delineate the CTV.}
\centering
\begin{tabular}{llllll}
\hline
 & Threshold: & Threshold: & \textbf{Threshold:} & Threshold: & Threshold:\\
Metric & level 1 & level 2 & \textbf{level 3} & level 4 & level 5\\
\hline
Dice (\%) & 83.50 (7.16) & 85.69 (6.32) & \textbf{86.76 (5.39)} & 86.15 (5.48) & 78.78 (12.66)\\
Accuracy (\%) & 93.46 (2.93) & 95.10 (2.15) & \textbf{96.03 (1.36)} & 96.27 (1.01) & 95.65 (1.99)\\
Sensitivity (\%) & 81.54 (9.45) & 84.58 (9.18) & \textbf{85.98 (9.81)} & 84.82 (11.48) & 72.49 (20.32)\\
Specificity (\%) & 96.72 (3.31) & 97.38 (2.41) & \textbf{97.90 (1.72)} & 98.19 (1.37) & 99.06 (0.77)\\
HD95 (mm) & 25.00 (13.59) & 19.36 (13.20) & \textbf{16.43 (13.26)} & 16.09 (13.12) & 13.38 (4.64)\\
\hline
\end{tabular}
\end{table*}

\subsection*{Comparison to consensus GTV}
\label{sec:org7866387}

Confidence maps for the evaluation dataset obtained with the proposed deep
neural network and from the multiple human readers were compared to the
consensus GTV, established by all the readers in a joint contouring session.
When comparing a (discretized) confidence map with the consensus GTV, a
threshold on the confidence level is first selected to convert the confidence
map to a binary GTV mask.

\begin{figure}[htbp]
\centering
\includegraphics{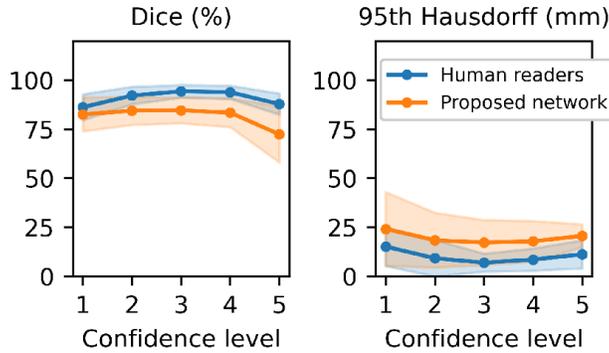}
\caption{\label{fig-cons-met-thr}Comparison between thresholded confidence maps and consensus GTV contours.  For each metric, both the human and predicted GTV confidence maps are thresholded and the resulting mask is compared to the consensus GTV using traditional binary metrics.  Plots show the metric as a function of the threshold.  Shaded areas correspond to one standard deviation around the mean.  The level of agreement with the consensus GTV achieved by the proposed method approaches that of the human confidence map.}
\end{figure}

The resulting metrics for varying thresholds are compared in
Figure~\ref{fig-cons-met-thr}.  The figure shows the Dice, sensitivity,
specificity and 95th percentile Hausdorff distance between thresholded
confidence maps (from human readers and the proposed neural network) and the
consensus GTV.  The range of metric values largely overlap which suggests that
the proposed network generated reasonable confidence maps.  While human GTVs
tended to better match the consensus, Dice scores between thresholded predicted
GTV and the consensus GTV were around 84.6\% (\(\pm 6.6 \%\)).  This difference can
be explained by the fact that consensus GTVs were drawn by the same human
readers who performed the multiple GTV delineations, therefore introducing an
unavoidable bias.  The 95th percentile Hausdorff distance to the consensus GTV
for predicted GTVs was typically within 10 mm of that of human GTVs.  This
confirms that the proposed network calculated GTV masks that are comparable to
masks from different readers.

\subsection*{CTV comparison}
\label{sec:org0551eb3}

\begin{figure*}[!htb]
  \centering
  \includegraphics{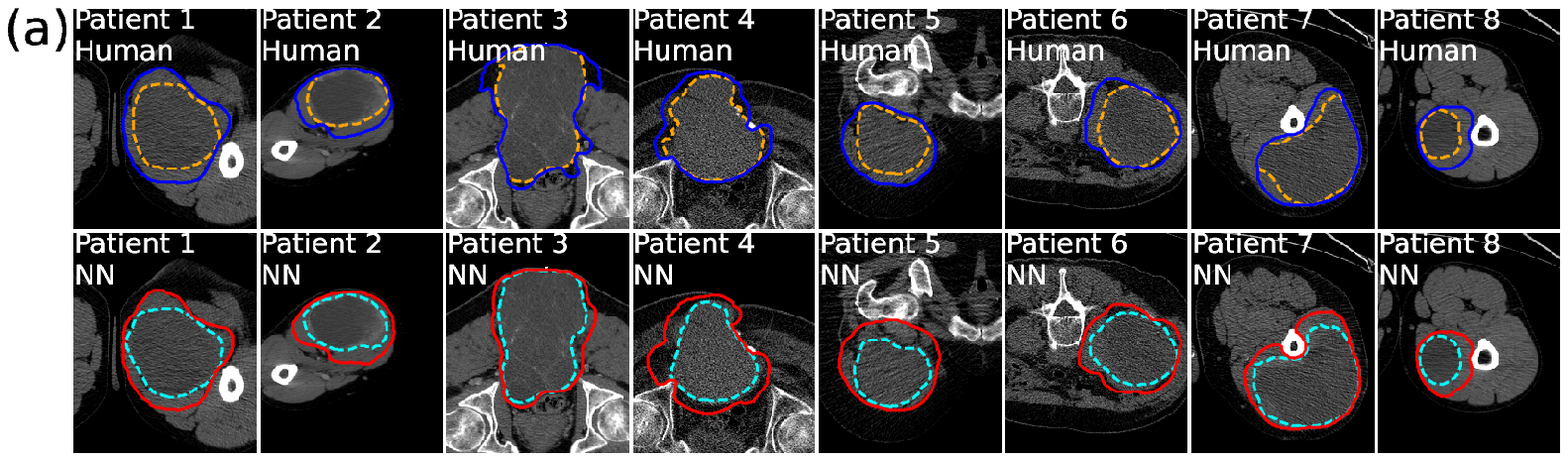}
  \\
  \includegraphics{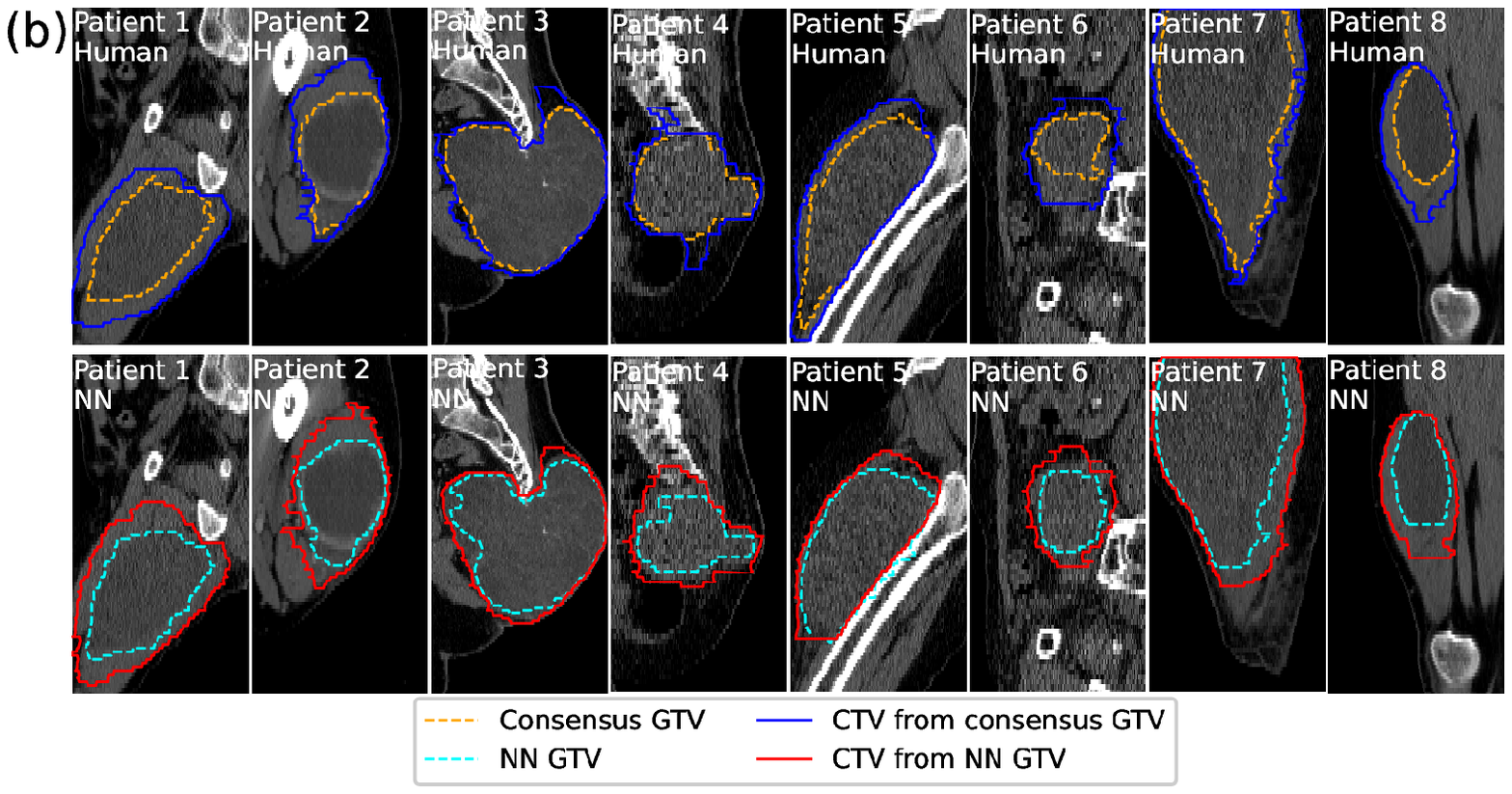}
  \caption{\label{fig-ctv-res}CTV comparison.  Comparison of CTVs drawn from the
    human consensus GTV and from GTV predicted by the proposed neural network
    (thresholding predicted confidence maps at confidence level 3).  The figure
    demonstrates a good agreement between CTVs obtained from human and the
    proposed neural network.}
\end{figure*}

The human consensus GTV and predicted GTV obtained by thresholding the predicted
confidence map for each patient in the evaluation were used by a human reader to
determine the CTV.  The CTV was drawn according to current standard clinical
practice which includes a 3 cm longitudinal expansion into tissues of least
resistance and 1 to 1.5 cm margin radially across tissues of greater resistance.
The resulting CTV contours are shown, along with the corresponding GTV contours
in Fig.~\ref{fig-ctv-res}.  The agreement (Dice score) between CTVs
derived from human consensus and predicted GTVs was around 89.5\% (\(\pm 1.8 \%\)),
which is even higher than the agreement between the corresponding GTVs (84.6\%
\(\pm 6.6 \%\)).

\section*{Discussion}
\label{sec:orge57b534}

We have demonstrated that the proposed deep learning method can accurately
predict the GTV for soft-tissue and bone sarcomas from CT images.  A continuous
Dice score of 86.7\% was achieved between predicted and human confidence maps and
a binary Dice score of 84.6\% was achieved between the human consensus GTV and
the GTV obtained by thresholding the predicted confidence maps.  We have shown
that the agreement with the consensus GTV approaches that of typical human
readers.

The proposed method offers multiple opportunities to streamline radiation
therapy.  It can be used as an initial guide for GTV contouring and allow human
readers to focus their time on delineation of regions with lower confidence
levels, substantially improving efficiency.  Such a tool can also be valuable to
assist less experienced readers or when readers perform GTV contouring of
targets outside their specialty.  Furthermore, automatic prediction of GTV
contours can enable adaptive planning by allowing GTV contours to be estimated
in a fraction of the time required by human readers.  The confidence maps, when
coupled with PET or MRI information can also be used to modulate dose and select
candidate regions for radiation boost based on the confidence level, which would
lead to better tumor control.

Another advantage of the training strategy using discretized confidence maps is
that it can be used with any number of contours per patient, reducing to
traditional binary segmentation task if only one contour is available.  The
availability of public datasets (typically with one GTV per image) can then
allow pre-training the network, before using transfer learning to refine network
weights with additional datasets (that may contain multiple contours for each
patient).

Future work will integrate other modalities (MRI and PET) in the automatic GTV
estimation network.  We expect that the inclusion of additional imaging
modalities will improve the training capabilities, resulting in even better
agreement with human readers for GTV and CTV.

\section*{Conclusions}
\label{sec:orgd4c3887}

We have proposed and validated an automatic GTV segmentation method, modeling
inter- and intra-reader variability and predicting GTV confidence maps.  We
demonstrated the performance of the proposed method by comparing the predicted
confidence maps to confidence maps obtained from human readers and to a
consensus GTV established jointly by all readers.  For both comparisons, we have
shown that the agreement between the predicted GTV and the reference is in the
same range as the typical agreement observed between readers.

Therefore, the proposed deep learning method demonstrated promise in automating
gross target volume delineation with performance comparable to that of a
consensus clinical human observer.

\section*{Acknowledgments}
\label{sec:org062abad}

This work was supported in part by the National Institutes of Health under
awards: T32EB013180, R01CA165221 and P41EB022544.

\bibliographystyle{model1-num-names}
\bibliography{bibliography}

\pagebreak
\onecolumn

\section*{Supplementary material}
\label{sec:orgb7dca67}

\renewcommand{\thepage}{S\arabic{page}}
\renewcommand{\thesection}{S\arabic{section}}
\renewcommand{\thetable}{S\arabic{table}}
\renewcommand{\thefigure}{S\arabic{figure}}
\renewcommand{\figurename}{Supplemental Material, Figure}
\renewcommand{\tablename}{Supplemental Material, Table}
\setcounter{figure}{1}
\setcounter{table}{0}
\setcounter{page}{1}

\subsection*{Supplementary material S1: Network parameters}
\label{sec:orgbb7f78b}

\begin{table}[htbp]
\caption{\label{table-sup-network-params}List of parameters for convolutional neural network training.}
\centering
\begin{tabular}{lr}
\hline
Network parameter & Value\\
\hline
Input image size & [256, 256, 5]\\
Number of trainable parameters & 2097946\\
Number of training steps & 100000\\
Optimizer & ADAM \cite{Kingma2015}\\
Learning rate & 0.008819\\
Batch size & 16\\
Cross-validation: \# of folds & 4\\
Cross-validation: search method & Tree of Parzen Estimators (TPE) \cite{Bergstra2011}\\
Class weights (\(w_c\)) & [0.000589, 0.238029, 0.266232, 0.296429, 0.139155, 0.059564]\\
Dropout & 0.140508\\
Weight \(\ell_2\) regularization & $2.754444\times10^{-9}$\\
Post-processing (inference) & 2-voxel morphological opening then closing +\\
 & small regions removal (\(< 50\) voxels)\\
\hline
\end{tabular}
\end{table}
\end{document}